\documentclass[pra, showpacs, twocolumn]{revtex4}

\usepackage{graphicx}
\usepackage{amsmath}

\begin{document}

\newcommand{ {\JOneZero} }{{${J_i = 1} \leftrightarrow {J_f = 0}~$}}
\newcommand{ \SPD } {{${^2 S_{1/2}} \leftrightarrow {^2 P_{1/2}} \leftrightarrow {^2 D_{3/2}}~$}}
\newcommand{ \SP }{{${^2 S_{1/2}} \leftrightarrow {^2P_{1/2}}~$}}
\newcommand{ \DP }{{${^2 D_{3/2}} \leftrightarrow {^2 P_{1/2}}~$}}

\title{Destabilization of dark states and optical spectroscopy in Zeeman-degenerate atomic systems}

\author{D. J. Berkeland}
\email{djb@lanl.gov}
\affiliation{Los Alamos National Laboratory, Physics Division, P-23, MS H803, Los Alamos, NM 87545}
\author{M. G. Boshier}
\email{m.g.boshier@sussex.ac.uk}
\affiliation{Sussex Centre for Optical and Atomic Physics, University of Sussex, Brighton BN1 9QH, UK}

\begin{abstract}
We present a general discussion of the techniques of destabilizing
dark states in laser-driven atoms with either a magnetic field or
modulated laser polarization. We show that the photon scattering
rate is maximized at a particular evolution rate of the dark
state.  We also find that the atomic resonance curve is
significantly broadened when the evolution rate is far from this
optimum value.  These results are illustrated with detailed
examples of destabilizing dark states in some commonly-trapped
ions and supported by insights derived from numerical calculations
and simple theoretical models.
\end{abstract}

\pacs{32.60.+i  42.50.Hz}
\maketitle

\section{Introduction}
\label{sec:Intro}

Under certain conditions a laser-driven atom will optically pump
into a dark state, in which the atom cannot fluoresce because the
coupling to every excited state vanishes
\cite{Orriols:1979,Arimondo:1976}. When the dark state is an
angular momentum eigenstate, this phenomenon is called optical
pumping. In the more general case in which the dark state is a
coherent superposition of angular momentum eigenstates and the
matrix elements vanish through quantum interference, it is usually
called ``coherent population trapping'' \cite{Gray:1978}. Dark
states are the basis of the STIRAP \cite{For:1998} method for
adiabatic manipulation of atomic states and the VSCPT
\cite{Aspect:1988} sub-recoil laser cooling scheme, and they are
closely related to the quantum interference-based mechanism of
lasing without inversion \cite{Harris:1989}. However, sometimes
population trapping in dark states is detrimental rather than
beneficial, such as when fluorescence is used to determine the
state of an atom \cite{obald:1989} or when atoms are laser-cooled
by scattering photons \cite{Wineland:1975, Hansch:1975}. Trapped
ions illustrate this point very well, because their superb
isolation from external perturbations can make the lifetime of
coherent dark states extremely long. In such systems, laser
cooling rates will be drastically reduced when the atom pumps into
a dark state.  One also finds that the laser cooling transition
can be significantly broadened when there is significant
accumulation of population in dark states, thereby reducing the
cooling rate. Although some methods for restoring atomic
fluorescence have been briefly discussed in the context of
specific atoms, a general prescription for destabilizing dark
states does not seem to exist. The aim of this paper is to provide
this information, supported by insights derived from numerical
calculations and simple theoretical models.

The paper is laid out as follows. In Sec.~\ref{sec:DarkStates} we
review the physics of dark states \cite{Arimondo:1996,
Bruce:1990}. In Sec.~\ref{sec:Destab} we discuss in general terms
some methods of destabilizing dark states. These techniques
involve either splitting the atomic energy levels (most commonly
with a magnetic field \cite{Janik:1985, Janik:1984}), or
modulating the laser polarization \cite{Jon:1994, Barwood:1998,
Berkeland:1998}. In Sec.~\ref{sec:DestabSpecific} we consider the
application of these methods to the level systems encountered in
some commonly-trapped ions. There we solve the equations of motion
for the atomic density matrix to find the excited state population
as a function of the various experimental parameters. We also
discuss destabilizing dark states in generic two-level atoms. We
conclude in Sec.~\ref{sec:Discussion} with a brief discussion of
the implications of these calculations and a summary of the
general prescription for destabilizing dark states.

\section{Dark states}
\label{sec:DarkStates}

We will mostly consider a two-level atom in which the total
angular momenta of the lower and upper levels are $J_i$ and $J_f$
respectively, with corresponding magnetic quantum numbers $m_i$
and $m_f$. In many cases of interest the atomic levels are further
split by the hyperfine interaction, which introduces the
possibility of optical pumping into other lower state hyperfine
levels. We will address this complication briefly in
Sec.~\ref{subsec:HigherAM}. Since we will be primarily interested
in scattering photons at a high rate we consider only electric
dipole transitions.

The atom is driven by a laser field
\begin{equation}
 \textsf{\textbf{E}}(t) = \frac{1}{2} \textbf{E}(t) e^{-i \omega_L t} + c.c.~,
\end{equation}
where $c.c.$ denotes the complex conjugate and $\omega_L$ is the
laser frequency.  It is convenient to expand the slowly-varying
amplitude $\textbf{E}(t)$ in terms of its irreducible spherical
components. These are related to the Cartesian components by
\cite{Weissbluth:1978}, \cite{Edmonds:1960}
\begin{subequations}
\begin{eqnarray}
 E_{\pm 1} &=& \mp \frac{1}{\sqrt {2}}(E_x \pm i E_y) \\
 E_0 &=& E_z.
 \end{eqnarray}
 \end{subequations}
The advantage of this decomposition is that the components
$E_{+1}$, $E_{0}$, and $E_{-1}$ (corresponding to $\sigma^{-}$,
$\pi$, and $\sigma^{+}$ polarizations) drive pure $\Delta m = -1,
0, \text{~and~} + 1$ transitions respectively. In the the rotating
wave interaction picture \cite{Bruce:1990}, the electric dipole
interaction operator $- e \textbf{r} \cdot \textsf{\textbf{E}}(t)$
is then
\begin{equation}
 - \frac{1}{2} e \textbf{r} \cdot \textbf{E} = \frac{1}{2} e r \left( E_{+1} C_{-1}^{(1)} -
E_{0} C_{0}^{(1)} + E_{-1} C_{+1}^{(1)} \right),
\end{equation}
where $C_{q}^{(1)} \equiv \sqrt {4\pi} Y_{1q} $ are the components
of the reduced spherical harmonic of rank 1. The non-zero matrix
elements of this operator are given by the Wigner-Eckart theorem
as
\begin{eqnarray}
\label{eq:WignerEckart}
 &{}& \left\langle J_i, m_i \left| - \frac{1}{2}e \textbf{r} \cdot \textbf{E} \right| J_f, m_f \right\rangle
 \nonumber\\
 &=&  - \frac{1}{2}(-)^{J_i - m_i}
 \nonumber \\
 &&\times \sum_{q = - 1}^{1} {( -
)^{q}E_{-q} \left( \begin{array}{ccc}
 J_i & 1 & J_f\\
-m_i & q & m_f\\
\end{array} \right)
\left\langle J_i \left \| e r \right \| J_f \right\rangle}.
\end{eqnarray}
The transition Rabi frequency $\Omega_{m_i m_f}$ is defined, as
usual, so that the matrix element (\ref{eq:WignerEckart}) is equal
to $-\frac{1}{2}\hbar \Omega_{m_i m_f}$. We will also make
frequent use of the rms Rabi frequency $\Omega$ \cite{Bruce:1990},
defined by the relationship
\begin{equation}
\label{eq:rmsRabi} \Omega^2 = \sum_{m_i, m_f} \left|\Omega_{m_i
m_f} \right| ^ 2
\end{equation}

Now, a general superposition of lower level states
\begin{equation}
\label{eq:genDark} \left| d \right\rangle = \sum_{m_i} {c_{m_i}
\left| J_i , m_i \right\rangle}
\end{equation}
will be dark if the electric dipole matrix element
\begin{equation}
\label{eq:dDotE}
\left\langle d \left| -e \textbf{r} \cdot
\textbf{E} \right| f \right\rangle = 0
\end{equation}
vanishes for every excited state $| f \rangle$. For example, in
the simple \JOneZero system, Eqs.~(\ref{eq:genDark}) and
(\ref{eq:dDotE}) give a single equation for the dark state
\begin{equation}
\label{eq:dDotE10}
 c_{-1} E_{+1} + c_{0} E_{0} + c_{+1} E_{-1} = 0.
\end{equation}
This equation has a non-trivial solution for \textit{any} static
laser field \textbf{E}, which has the important consequence that
an atom driven on a \JOneZero transition by light of constant
polarization will always be pumped into a dark state. For this
transition, Eq.~(\ref{eq:dDotE10}) has a two-dimensional solution
space spanned by the states
\begin{subequations}
\begin{equation}
\left[\begin{array}{c}
 c_{-1} \\
 c_{0}  \\
 c_{+1} \\
\end{array} \right] = \frac{1}{\sqrt{E_{-1}^2 + E_{+1}^2}} \left[ \begin{array}{c}
 -E_{-1} \\
 0       \\
 E_{+1} \\
\end{array} \right]
\end{equation}
and
\begin{eqnarray}
\left[ \begin{array}{c}
 c_{-1} \\
 c_{0} \\
 c_{+1} \\
\end{array}  \right]  = \frac{1}{\sqrt{(E_{-1}^2 + E_{+1}^2)(E_{-1}^2 + E_{0}^2 +
E_{+1}^2)}} \nonumber\\
 \times \left[\begin{array}{c}
 -E_{0} E_{+1} \\
 E_{-1}^2 + E_{+1}^2 \\
 E_{0} E_{-1}\\
\end{array}  \right],
\end{eqnarray}
except when the laser light is $\pi$-polarized, in which case the
space of dark states is spanned by
\begin{equation}
 {\left[\begin{array}{c}
 c_{-1} \\
 c_{0}  \\
 c_{+1} \\
\end{array} \right] = \left[ \begin{array}{c}
 1 \\
 0 \\
 0 \\
\end{array}\right]} \textrm{~and~}
 {\left[\begin{array}{c}
 c_{-1} \\
 c_{0}  \\
 c_{+1} \\
\end{array} \right] = \left[ \begin{array}{c}
 0 \\
 0 \\
 1 \\
\end{array}\right]}
\end{equation}

\end{subequations}

Table~\ref{tab:darkTable1} summarizes the conditions under which
Zeeman degenerate systems can have dark states.  It shows that
there is always at least one dark state if $J_f=J_i-1$, or if
$J_f=J_i$ and $J_f$ is an integer. These are the cases which we
address in this paper.  The dark states for the simplest of these
systems, found by solving Eq.~(\ref{eq:dDotE}), are listed in
Table~\ref{tab:darkTable2}.

\begin{table}
\caption{Existence of dark states for arbitrary atomic systems and
laser polarization in zero magnetic field.} \label{tab:darkTable1}
\begin{ruledtabular}
\begin{tabular}{c p{1.2in}p{1.2in}}
{Upper level}& \multicolumn{2}{c}{Lower level}\\
 {$J_f$}& Integer $J_i$ & Half-integer $J_i$  \\
\hline
$J_i$ + 1 & No dark state & No dark state \\
$J_i$ & One dark state for any polarization&One dark state for circular polarization only \\
$J_i-1$& Two dark states for any polarization&Two dark states for any polarization \\
\end{tabular}
\end{ruledtabular}
\end{table}

\begingroup
\begin{table*} \caption{Unnormalized dark states for
several atomic systems when laser light is not $\pi$-polarized.
For $\pi$-polarized light, the dark states are $\vert J_i, m_i=0
\rangle$ for integer $J_i \leftrightarrow J_i$ transitions and
$\vert J_i, m_i=\pm J_i \rangle$ for $J_i \leftrightarrow J_i-1$
transitions} \label{tab:darkTable2}
\begin{ruledtabular}
\begin{tabular}{ccc}
$J_i$ & $J_f$ & dark states(s) $ \left[ \begin{array}{c} c_{ - J_i} \\
\vdots
\\c_{ + J_i} \end{array} \right] $
\vspace{2pt} \\
\hline \\
1 & 0 &
      $ {\left[ \begin{array}{c} {E_{-1}} \\ {0} \\ {E_{+1}} \end{array}  \right]} $
        and
      $ {\left[ \begin{array}{c} {E_{0} E_{+1}} \\ { - \left( {E_{-1}^2 + E_{+1}^2 }  \right)}
          \\ {E_{0} E_{-1}} \\ \end{array}  \right]} $
\\ \\
1 & 1 &
    $ {\left[ {{\begin{array}{c}{E_{-1}}\\{ - E_{0}}\\ {E_{+ 1}} \\ \end{array}}}  \right]}$
\\ \\
$\frac{3}{2}$ & $\frac{1}{2}$ &
$ {\left[ {{\begin{array}{c}
 {\sqrt {2} E_{-1} E_{0}}\\
 { - \sqrt {3} E_{-1} E_{+1}}\\
 {0}\\
 {E_{+1}^2 }\\
\end{array}}}  \right]}
$
and
  $
{\left[ {{\begin{array}{c}
 {E_{+1} \left( {3E_{-1}^2 + E_{-1} E_{+1}^2 - 2E_{0}^2 E_{+1}}  \right)}\\
 {\sqrt {6} E_{0} \left( {E_{-1}^{3} + E_{+1}^{3}}  \right)} \\
 { - \sqrt {3} \left( {E_{+1}^{4} + E_{-1}^2 \left( {2E_{0}^2 + 3E_{+1}^2 }  \right)} \right)}\\
 { - \sqrt {2} E_{-1} E_{0} \left( {E_{+1} \left( {E_{-1} - 3E_{+1}}  \right) - 2E_{0}^2 }  \right)}\\
\end{array}}}  \right]}
$
\\ \\
2 & 1 &
$ {\left[ {{\begin{array}{c}
 {E_{-1} \left( {E_{-1} E_{+1} - 2E_{0}^2 }  \right)} \\
 {\sqrt {8} E_{-1} E_{0} E_{+1}}  \\
 { - \sqrt {6} E_{-1} E_{+1}^2 }  \\
 {0} \\
 {E_{+1}^{3}}\\
\end{array}}}  \right]}
$
and
$ {\left[ {{\begin{array}{c}
 { - \sqrt {2} E_{0} E_{+1} \left( {2E_{0}^2 + 3E_{-1}^{3} E_{+1} + E_{-1} E_{+1}^{3} - E_{0}^2 E_{+1}^2 }  \right)}\\
 {E_{-1} E_{+1} \left( {E_{-1}^{4} + 6E_{-1}^2 E_{+1}^2 + E_{+1}^{4}}  \right) - 2E_{0}^2 \left( {E_{-1}^{4} + E_{+1}^{4}}  \right)} \\
 { - \sqrt {3} E_{0} \left( { - E_{-1}^{4} E_{+1} + E_{+1}^{5} + 2E_{-1}^{3} \left( {2E_{+1}^2 + E_{0}^2 }  \right)} \right)} \\
 { - E_{+1}^2 \left( {E_{-1}^{4} + 6E_{-1}^2 E_{+1}^2 + E_{+1}^{4}}  \right) + 4E_{-1}^2 E_{0}^2 E_{+1} \left( {E_{-1} - 2E_{+1}}  \right) - 4E_{-1}^2 E_{0}^{4}} \\
 {\sqrt {2} E_{-1} E_{0} \left( {E_{+1}^2 \left( {E_{-1}^2 - 2E_{-1} E_{+1} + 3E_{+1}^2 }  \right) + E_{0}^2 E_{+1} \left( { - 3E_{-1} + 4E_{+1}}  \right) + 2E_{0}^{4}}  \right)} \\
\end{array}}}  \right]}
$
\\ \\
2&2&
$
{\left[ {{\begin{array}{c}
 {E_{-1}^2 }\\
 { - \sqrt {2} E_{-1} E_{0}}\\
 { - \sqrt{\frac{2}{3}} \left( {E_{-1} E_{+1} + E_{0}^2 }  \right)}\\
 { - \sqrt {2} E_{0} E_{+1}} \\
 {E_{+1}^2 }
 \ \\
\end{array}}}  \right]}
$
\end{tabular}
\end{ruledtabular}
\end{table*}
\endgroup

\section{Destabilizing dark states: general principles}
\label{sec:Destab}

To decrease population accumulation in dark states by making them
time-dependent, one can either shift the energies
$\varepsilon_{m_i}$ of the states $\vert m_i \rangle$ by unequal
amounts with an external field, or modulate the polarization of
the laser field $\textbf{E}(t)$. The general instantaneous dark
state (\ref{eq:genDark}) then evolves in time as
\begin{eqnarray}
\label{eq:timeDepGenDark}
 \left|d(t)\right\rangle  =
\sum_{m_i} {c_{m_i} \left( \textbf{E}(t) \right) \left| J_i, m_i
\right\rangle e^{-i \varepsilon_{m_i} t / \hbar}},
\end{eqnarray}
where we have made explicit the dependence of the dark state
components $c_{m_i}$ on the laser field
(Table~\ref{tab:darkTable2}). The application of a magnetic field
is probably the simplest and most widely used method of
destabilizing dark states. But there are systems (for example,
trapped-ion frequency standards \cite{Alan:2001, Peter:1997}) in
which external fields cannot be tolerated.  In these cases the
laser field must be modulated instead.

We will evaluate the effectiveness of a destabilization technique
by calculating the excited state population (proportional, of
course, to the fluorescence rate) as a function of experimental
parameters. This is done computing the evolution of the atomic
density matrix $\rho$ using the Liouville equation of motion
\begin{equation}
\label{eq:Liouville}
 \frac{\partial \rho}{\partial t} =
\frac{1}{i\hbar} \left[ H , \rho \right] + \frac{\partial \rho
_{relax}}{\partial t},
\end{equation}
where $H$ includes both the rotating wave interaction picture
Hamiltonian for the coupling to the laser \cite{Bruce:1990} and
the Zeeman interaction with the external magnetic field. The last
term in Eq.~(\ref{eq:Liouville}) accounts for the spontaneous
decay of the excited states and its effect on the ground states,
including the decay of coherences between the excited states into
coherences between ground states \cite{Cardimona:1982},
\begin{eqnarray}
\label{eq:gendecay} &&\frac{\partial \rho_{m_i m_i'}}{\partial
t} \nonumber\\
&=&
(-)^{m_i-m_i'} (2J_f  + 1) \gamma \nonumber\\
&\times& \sum\limits_{m_f m_f'}{}
 {\left( {\begin{array}{ccc}
   {J_f } & {J_i } & 1  \\
{ - m_f } & {m_i } & q  \\
 \end{array}} \right)
\left( {\begin{array}{ccc}
   {J_f } & {J_i } & 1  \\
{- m_f '} & {m_i'} & q  \\
\end{array}} \right)
 \rho _{m_f m_f'}}\nonumber\\
\end{eqnarray}
where $q$ has the value which causes the 3-$j$ symbols to be
non-zero. Equation~(\ref{eq:Liouville}) results in a system of
$4(J_i+J_f+1)^2$ coupled differential equations which can be
solved numerically and, in some simple cases,
analytically~\cite{Many:1}. When the laser polarization is static,
the steady state solution is easily found by solving
Eq.~(\ref{eq:Liouville}) with ${\partial \rho}/{\partial t} = 0$.
When the laser field is modulated (with period $T$), the density
matrix will, in general, evolve towards a quasi-steady-state
solution in which $\rho(t)=\rho(t+T)$. In these cases we compute
the excited state population by averaging the quasi-steady state
solution over a modulation period.

At this point, the main result of this paper can already be
summarized qualitatively. There are always three relevant time
scales in the problem, with three corresponding frequencies: the
excited state decay rate $\gamma$, the resonant Rabi frequency
$\Omega$, and the laser polarization modulation frequency or
energy level shift $\delta$. The parameter $\delta$ characterizes
the evolution rate of the dark state (\ref{eq:timeDepGenDark}) (if
necessary, the exact time evolution of the dark state can be found
by substituting the shifted energies or the time-dependent laser
field and the appropriate dark state from
Table~\ref{tab:darkTable2} into Eq.~(\ref{eq:timeDepGenDark})). We
find from our simulations that the evolution rate which maximizes
the excited state population is $\delta \sim \Omega/2$. The
excited state population is never large when $\Omega$ and $\delta$
are significantly different:  it is small if $\Omega$ is too large
because then the strongly-driven atom is able to follow the
evolving dark state adiabatically, and also if $\delta$ is too
large because then the atom and the laser become detuned. We find
also that the transition line shape is broadened in both of these
limits ($\Omega / \delta \gg 1$ and $\Omega / \delta \ll 1$). This
can be a practical concern when laser cooling an ion, because the
Doppler-limited temperature of a Doppler-cooled atom is
proportional to the resonance width \cite{Wayne:1982}.  The
ultimate temperature of the ion can therefore be substantially
increased if $\delta$ is not optimum, both because the scattering
rate is decreased and also because the width of the transition is
increased.  Fortunately, the conditions that maximize the excited
state population also minimize the transition linewidth, and we
find that making both $\Omega$ and $\delta$ about a fifth of the
decay rate $\gamma$ leads to appreciable excited state population
without significantly broadening the transition.

\section{Destabilizing dark states in specific atomic systems}
\label{sec:DestabSpecific}

We now consider the application of the techniques discussed above
to some commonly used atomic systems. We begin in
Sec.~\ref{subsec:J10} with the \JOneZero system, which illustrates
the basic properties of the destabilization methods in two-level
systems. Then, in Sec.~\ref{subsec:SPD}, we discuss the
bichromatic $\Lambda$-system of ${J_{i1}=\frac{1}{2}}
\leftrightarrow {J_f=\frac{1}{2}} \leftrightarrow
{J_{i2}=\frac{3}{2}}$. This complex system illustrates several
important consequences of dark states for some commonly-trapped
ions. Finally, in Sec.~\ref{subsec:HigherAM} we generalize these
results to two-level systems with higher values of the total
angular momentum (the specific case of the $J=5 \leftrightarrow
J=5$ system has already been discussed elsewhere by one of us
\cite{Boshier:2000}).

When a magnetic field \textbf{B} is applied to the atom, we will
choose the quantization axis to be parallel to \textbf{B}, and we
will assume that the laser light is linearly polarized at an angle
$\theta_{\text{BE}}$ to \textbf{B}. This choice of polarization
makes the calculations somewhat simpler, and it is generally more
straightforward to implement in the laboratory than solutions
using circularly or elliptically polarized light. Also, if a
transition has a dark state, we find that driving it with
circularly or elliptically polarized light does not significantly
change the optimum efficiency of the techniques discussed here. We
will measure the strength of the magnetic field in terms of the
Zeeman shift frequency $\delta_B=\mu_B |\textbf{B}| / \hbar$,
where $\mu_B$ is the Bohr magneton. The detuning of the laser
frequency $\omega_L$ from the unperturbed atomic resonance
frequency $\omega_0$ is $\Delta = \omega_L - \omega_0$, and the
total decay rate of the excited level is $\gamma$. Finally, we
will usually assume that laser linewidths are negligible compared
to the relevant decay rates.

\subsection{ { \JOneZero } }
\label{subsec:J10}

The simple ${^2 S_{1}} \leftrightarrow {^2 P_{0}}$, \JOneZero
transition is well-suited to discussing in detail the main methods
of destabilizing dark states.  Also, the closely-related nuclear
spin $I=\frac{1}{2}$, $F_i=1 \leftrightarrow F_f=0$ transition is
used in laser cooling and fluorescence detection in both
$^{199}\text{Hg}^+$ \cite{Berkeland:1998} and $^{171}\text{Yb}^+$
\cite{Roberts:1999, Enders:1993, Fisk:1997, Tamm:1995} ($F$
denotes as usual the total angular momentum in atoms with non-zero
nuclear spin). Decays to the $F_i=0$ state in these systems are
infrequent, occurring only through off-resonance excitation of the
$F_f=1$ level. The atom can be pumped out of the $F_i=0$ state by
a microwave field driving the ${\vert F_i=0,m_i=0\rangle}
\leftrightarrow {\vert F_i=1,m_i=0 \rangle}$ transition
\cite{Roberts:1999, Enders:1993}, or by a laser field driving a
${\vert F_i=0,m_i=0\rangle} \leftrightarrow {\vert F_f=1\rangle}$
transition \cite{Berkeland:1998}. As long as this pumping mostly
keeps the population within the $F_i=1$ and $F_f=0$ levels, the
results for these real systems are nearly identical to those for
the $I=0$, ${^2 S_{1}} \leftrightarrow {^2 P_{0}}$ transition we
discuss here.

The next two sections discuss destabilizing dark states in this
system, firstly with a magnetic field, and then with polarization
modulation.

\subsubsection{Destabilization with a magnetic field}
\label{subsubsec:J10BField}

The steady-state solution of Eq.~(\ref{eq:Liouville}) for the
\JOneZero system in a magnetic field can be found analytically.
The resulting expression for the excited state population is
\begin{equation}
\label{eq:j10Pop}
 P_f = \frac{3}{4} \frac{\Omega ^2 \cos^2 \theta_{\text{BE}} \sin^2\theta_{\text{BE}}}
 {1 + 3\cos^2\theta_{\text{BE}}} \frac{1}{(\gamma'/2)^2 + \Delta ^2},
\end{equation}
where
\begin{eqnarray}
\label{eq:j10Width}
  \left( \frac{\gamma'}{2}\right)^2
= \left( \frac{\gamma}{2}\right)^2 + \Omega ^2 \cos^2
\theta_{\text{BE}} \frac{1 - 3 \cos^2
\theta_{\text{BE}}}{1 + 3 \cos^2\theta_{\text{BE}}} \nonumber\\
+ \frac{\cos^2\theta_{\text{BE}}}{1 + 3\cos^2\theta_{\text{BE}}}
\left( \frac{\Omega^4}{16 \delta_B^2} + 16 \delta_{B}^2 \right),
\end{eqnarray}
and $\Omega$ is the rms Rabi frequency defined in
Eq.~(\ref{eq:rmsRabi}).  In our normalized units, the Zeeman
shifts of the $m_i=\pm 1$ ground states are $\pm 2 \delta_B$
respectively, so the dark state evolution rate is $\delta =
2\delta_B$.

\begin{figure}
\includegraphics{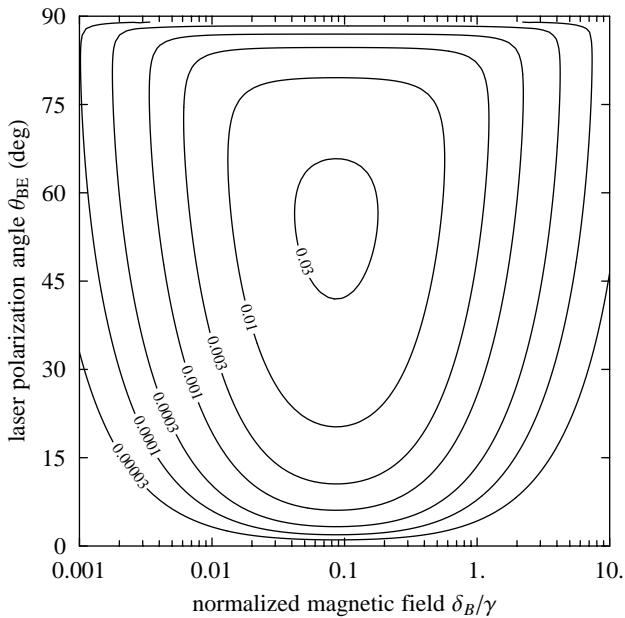}
\caption{Excited state population $P_f$ in the \JOneZero
transition as a function of normalized magnetic field $\delta_{B}
/ \gamma$ and laser polarization angle $\theta_{\text{BE}}$.  The
rms Rabi frequency $\Omega = \frac{\sqrt{3}}{5} \gamma$ and the
laser detuning $\Delta=0$.} \label{fig:j10vsAngle}
\end{figure}

\begin{figure}
\includegraphics{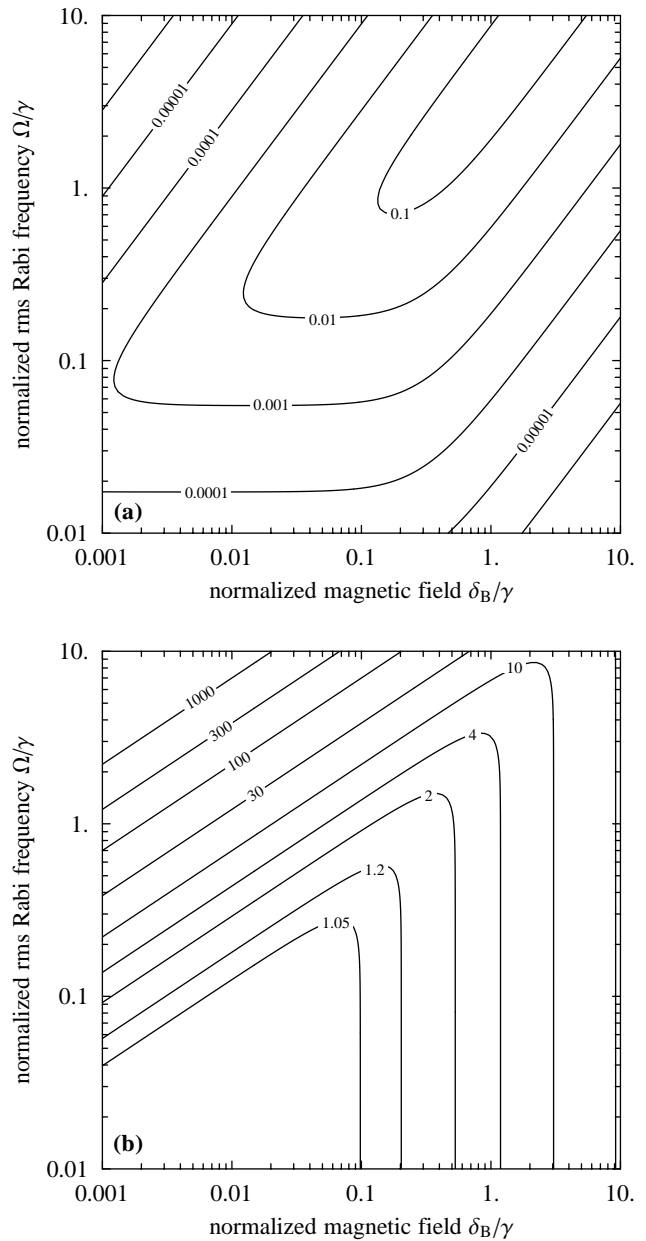}
\caption{(a) Excited state population, and (b) resonance width (in
units of $\gamma$) as a function of normalized magnetic field and
normalized rms Rabi frequency for the \JOneZero transition with
$\theta_{\text{BE}} = \arccos \frac{1}{\sqrt{3}}$ and $\Delta =
0$.} \label{fig:j10WidthAndHeight}
\end{figure}

Figure~\ref{fig:j10vsAngle} is a graph of the excited state
population $P_f$ as a function of magnetic field strength and
polarization angle for a convenient value of Rabi frequency.  It
shows that there is both an optimum magnetic field strength and an
optimum polarization angle for the laser field. $P_f$ vanishes
when $\theta_{\text{BE}}=0^{\circ}$ or $90^{\circ}$ because the
atom then optically pumps into the $m_i=\pm 1$ states or the
$m_i=0$ state respectively. One finds from Eq.~(\ref{eq:j10Pop})
that the excited state population is maximized for a given Rabi
frequency by choosing the magnetic field strength so that
$\delta_B = \Omega/4$ and the laser polarization angle so that
$\theta_{\text{BE}} = \arccos \frac{1}{\sqrt{3}}$ (the angle which
makes the three transition Rabi frequencies equal). Figure
\ref{fig:j10WidthAndHeight}(a) shows $P_f$ as a function of Rabi
frequency and magnetic field for this optimum angle.  The excited
state population is, as expected, small in both the low intensity
regime $\Omega < \gamma$ and in the large-detuning regime
$\delta_B > \gamma$.  However, it is also small even at high
intensity ($\Omega > \gamma$) if $\delta_B \ll \Omega$.  Our
simulations show that this occurs because under these conditions
the atom adiabatically follows the evolving instantaneous dark
state. On the other hand, near the optimum condition $\delta_{B} =
\Omega/4$ the dark state evolves quickly enough that the atom
never entirely pumps into it, and so the excited state population
can be substantial.

In many applications the linewidth of the transition is also an
important quantity. Fig.~\ref{fig:j10WidthAndHeight}(b) shows the
dependence of the resonance width $\gamma'$ on the Rabi frequency
and the magnetic field. We see that the linewidth is large when
$\delta_{B} > \gamma$ because of Zeeman broadening, when $\Omega >
\gamma$ because of ordinary power broadening (the second term in
Eq.~(\ref{eq:j10Width})), and also when $\delta_{B} \ll \Omega$, a
less obvious regime which will be discussed below.
Figures~\ref{fig:j10WidthAndHeight}(a) and
\ref{fig:j10WidthAndHeight}(b) illustrate the useful result
(easily obtained from Eqs. (\ref{eq:j10Pop}) and
(\ref{eq:j10Width})) that maximizing the excited state population
for a particular Rabi frequency also minimizes the resonance line
width.  These plots show that the choice $\Omega \sim \gamma/3$
gives substantial excited state population without significantly
increasing the linewidth.  If the linewidth is not important, then
the laser intensity can be chosen to saturate the transition
($\Omega \gg \gamma$) and the magnetic field strength then
adjusted so that $\delta_B = \Omega/4$.

\begin{figure}
\includegraphics{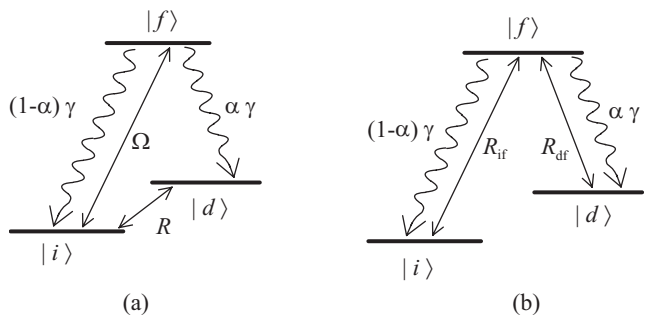}
\caption{Generic three-level $\Lambda $-systems with (a)
incoherent coupling between ground states, and (b) laser coupling
of both transitions.} \label{fig:3Lambda}
\end{figure}

The broadening of the resonance when $\delta_{B} \ll \Omega$ is
apparent in Eq.~(\ref{eq:j10Width}), which contains the terms
\begin{equation}
\label{eq:j10ExtraWidth}
 \frac{\cos^2 \theta_{\text{BE}}}{1 + 3 \cos ^2 \theta_{\text{BE}}}
\left( \frac{\Omega^4}{16 \delta_B^2} + 16 \delta_{B}^2 \right).
\end{equation}
The term proportional to $\delta_{B}^2 $ is simple Zeeman
broadening, but the term proportional to $1/\delta_{B}^2$ does not
have an obvious physical interpretation. We find that such a
broadening term is present whenever the atom contains a
$\Lambda$-system, regardless of the method of destabilizing the
dark state. To understand this behavior in simple physical terms,
consider Fig.~\ref{fig:3Lambda}(a), which shows a generic
$\Lambda$-system in which the laser field drives only one arm of
the system, with Rabi frequency $\Omega$. $|i\rangle$ and
$|d\rangle$ represent light and dark ground states respectively.
The excited state decays to the two ground states with branching
ratio $(1-\alpha):\alpha$. All of the systems discussed in this
paper can be described similarly (although with higher
multiplicity of states) after a change in basis states. For
example, in the \JOneZero transition driven by $\pi$-polarized
light, $|d\rangle$ corresponds to either the $m_i=1$ or $m_i=-1$
state, $|i\rangle$ corresponds to the $m_i=0$ ground state and
$|f\rangle$ corresponds to the excited state. For simplicity, in
this generic case the coupling between the light and dark ground
states is represented by an incoherent rate $R$, rather than the
coherent magnetic field. When the steady-state density matrix
equations are solved for this system, the population of the
excited state $|f\rangle$ is found to be
\begin{equation}
\label{eq:bottleneckPf}
 P_f =
\frac{\frac{1}{8} \Omega_{\text{if}}^2}{\frac{1}{4} \gamma^2 +
\frac{3}{8}\Omega_{\text{if}}^2 + \frac{1}{8}\alpha \gamma
\Omega_{\text{if}}^2 /R + \Delta_{\text{if}}^2},
\end{equation}
where $\Delta_{\text{if}}$ is the detuning of the laser frequency
from resonance with the atomic transition. The term $\frac{1}{8}
\alpha \gamma \Omega_{\text{if}}^2 / R$ is analogous to the
broadening term $\Omega ^{4} / 16 \delta_{B}^2 $ in
Eq.~(\ref{eq:j10Width}).  In both cases the linewidth becomes
large as the dark ground state is coupled less strongly to the
light ground state.

For trapped ions this behavior has a simple physical explanation
in terms of quantum jumps. Suppose that $\alpha \ll 1$, so most of
the decays are to state $|i\rangle$, and also that the pump rate
$R$ out of the dark state $|d\rangle$ is very small. Then, as long
as the atom is not in state $|d\rangle$, it behaves like a
two-level atom and we can use the well-known results for this
system. It follows that, on average, an atom initially in state
$|i\rangle$ will for a time $\Delta t = \frac{1 - \alpha}{\alpha}
(\frac{1}{4} \gamma ^2 +
\Delta_{\text{if}}^2)/\frac{1}{2}\Omega_{\text{if}}^2 \gamma$
scatter $(1-\alpha)/\alpha$ photons before decaying into state
$|d\rangle$. The atom will then not fluoresce for an average time
$1/R$, so that on average the number of fluorescent photons
emitted per unit time is
\begin{eqnarray}
\bar{n} &=& \frac{1-\alpha}{\alpha} \left(\frac{1}{R}+ \Delta t \right)^{-1} \nonumber\\
& \approx& \frac{1}{\alpha} \left( \frac{1}{R}+ \frac{1}{\alpha}
 \frac{\frac{1}{4}\gamma ^2 + \Delta_{\text{if}}^2}{\frac{1}{2} \Omega_{\text{if}}^2
 \gamma} \right)^{-1},
\end{eqnarray}
which agrees with the average photon emission rate $\gamma P_f
\sim R/\alpha$ obtained from Eq.~(\ref{eq:bottleneckPf}) in the
limit of small $R$. Now the source of the broadening can be seen:
if $R \ll \alpha \Omega_{\text{if}}^2 / \gamma$ and
$\Delta_{\text{if}} \lesssim \gamma$, then the time during which
the atom scatters photons is very short compared to the time that
it spends not fluorescing at all. The average rate of photon
scattering is therefore very small and does not change much with
detuning until $\Delta_{\text{if}}$ becomes much larger than
$\gamma$, because then the time it takes to pump into the dark
state is no longer less than the time the atom spends in the dark
state. When only the average photon scattering rate is measured,
the line shape as a function of laser frequency is therefore very
broad.  We will also consider below in
Sec.~\ref{subsubsec:SPDBField} an explanation of the broad line
shape at low incoherent pump rates in terms of rate equations
\cite{Janik:1985}.

\subsubsection{Destabilization with polarization modulation}
\label{subsubsec:J10PolMod}

We turn now to the second category of techniques for destabilizing
dark states, modulating the polarization state of the laser field.
In order to destabilize dark states in a \JOneZero system in zero
magnetic field, the three spherical components of the field
$E_{-1}$, $E_{0}$, and $E_{+1}$ must be non-zero and they must
have linearly-independent time dependences because otherwise
Eq.~(\ref{eq:dDotE10}) will have a nontrivial solution. Physically
this is because only one excited state is coupled to three ground
states, forming two conjoined $\Lambda $-systems that must be
independently destabilized. Imposing different time dependences on
all three polarization components requires two non-collinear laser
beams \cite{Berkeland:1998}.  This makes the \JOneZero system
experimentally more complicated than every other two-level system,
since Table~\ref{tab:darkTable2} shows that their dark states can
still be destabilized if one polarization component is zero and
the other two have different time dependences, which requires but
a single laser beam.

One obvious way of producing a suitable polarization modulation is
giving the three polarization components different frequencies.
This can be done, for example, by passing three linearly polarized
beams through separate acousto-optic modulators (AOM's), followed
by appropriately oriented waveplates. If right- and left-handed
circularly polarized light are separately shifted and copropagate
along the quantization axis, while a second beam is polarized
along the quantization axis, the resulting field can be written as
\begin{equation}
\left(\begin{array}{c}
 E_{-1} \\
 E_{0}  \\
 E_{+1} \\
\end{array} \right) =
\left( \begin{array}{c}
 E_{\sigma^+} e^{-i\delta_{\text{AOM-}} t }\\
 0       \\
 E_{\sigma^-} e^{-i\delta_{\text{AOM+}} t }\\
\end{array} \right) +
\left( \begin{array}{c}
 0     \\
 E_\pi   \\
 0     \\
\end{array} \right),
\end{equation}
where $\delta_{\text{AOM+}}$ and $\delta_{\text{AOM-}}$ are the
relative frequency shifts due to the AOM's, and $E_{\sigma^+}$,
$E_{\sigma^-}$, and $E_\pi$ are the amplitudes of the three laser
fields. The symmetrical conditions
$\delta_{\text{AOM+}}=-\delta_{\text{AOM-}}$ and
$E_{\sigma^+}=E_{\sigma^-}$ are most efficient at destabilizing
dark states.  In this case, the analytical solution to the density
matrix equation of motion (\ref{eq:Liouville}) is identical to
that obtained when a magnetic field with
$2\delta_B=\delta_{\text{AOM+}}$ is applied at an angle
$\theta_{\text{BE}} = \arctan(\sqrt{2} E_\pi / E_{\sigma^+})$, and
so the discussion accompanying Figs. \ref{fig:j10vsAngle} and
\ref{fig:j10WidthAndHeight} applies here also.  The optimum
parameters are therefore $|E_{+1}|=|E_{0}|=|E_{-1}|$ and
$\delta_{\text{AOM+}} = - \delta_{\text{AOM-}} = \Omega/2$ and
$\Omega \sim \gamma/3$.  Experimentally, this is perhaps the
simplest technique to destabilize dark states in this system,
because AOM's are simple, inexpensive devices.

An alternative technique for continuously altering the
polarization state of the field is modulating, with different
phases, the amplitudes of the polarization components of the
field. In atomic beam experiments, this can be done by sending the
atoms through two or more laser-atom interaction regions of
different laser polarization \cite{Jon:1994}. With trapped ions,
it has been demonstrated in several systems by smoothly varying
the intensity ratios or relative phases of the polarization
components of the light driving the stationary atoms
\cite{Barwood:1998, Berkeland:1998}. For example
\cite{Berkeland:1998}, overlapping at right angles a
$\pi$-polarized laser beam with a second beam that has passed
through a photo-elastic modulator (PEM) in which the fast axis is
compressed while the slow axis is expanded produces the field
\begin{equation}
\label{eq:PMfield} \left(\begin{array}{c}
 E_{-1} \\
 E_{0}  \\
 E_{+1} \\
\end{array} \right) =
\frac{E_{PEM}}{\sqrt{2}} \left( \begin{array}{c}
 e^{+i \varphi(t)} + i e^{-i \varphi(t)}\\
 0       \\
 e^{+i \varphi(t)} - i e^{-i \varphi(t)}\\
\end{array} \right) +
\left( \begin{array}{c}
 0     \\
 E_\pi   \\
 0     \\
\end{array} \right)
\end{equation}
where
\begin{equation}
\label{eq:PMphase}
\varphi(t) = \frac{1}{2} \Phi \left[ {1 - \cos
\left( {\delta_{PEM} t} \right)} \right],
\end{equation}
$\Phi$ is the phase modulation amplitude, and
$\delta_{\text{PEM}}$ is the modulation rate of the PEM indices of
refraction. As above, we have defined the quantization axis to be
parallel to the propagation direction of the modulated beam. If
$\Phi \ge \pi$, then the polarization of the modulated beam
continuously cycles between linear and right- or left-hand
circular polarizations. When $\Phi > \pi$, Fourier analysis of the
modulated field reveals a flat spectrum of harmonics of
$\delta_{\text{PEM}}$ up to a maximum harmonic number of $\sim
2\Phi/\pi$, so that the effective dark state evolution rate in
this high modulation index regime is $\delta \sim \Phi
\delta_{\text{PEM}}$.

\begin{figure}
\includegraphics{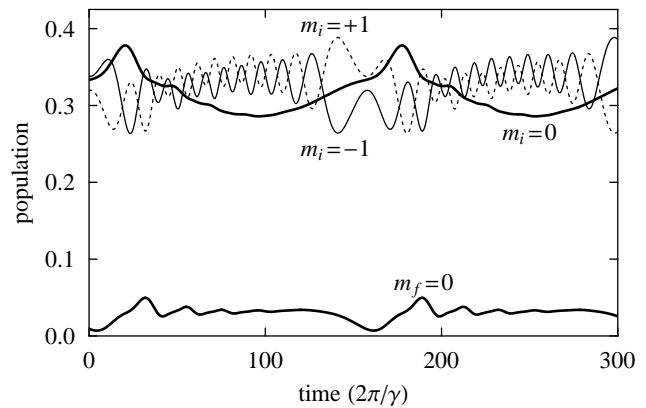}
\caption{Populations of individual states in the \JOneZero
transition when the laser field is modulated according to
Eq.~(\ref{eq:PMfield}) and the quasi-steady state has been
reached. The laser detuning $\Delta=0$, phase modulation amplitude
$\Phi=10\pi$, modulation frequency
$\delta_{\text{PEM}}=\gamma/50$, and Rabi frequencies
$\Omega_{-1,0}=\Omega_{0,0}=-\Omega_{+1,0}=\gamma/5$, so that the
rms Rabi frequency $\Omega = \frac{\sqrt{3}}{5}\gamma$, as in
Fig.~~\ref{fig:j10vsAngle}.} \label{fig:j10PolModVsTime}
\end{figure}

In this case the system never reaches a steady state. This can be
seen in Fig.~\ref{fig:j10PolModVsTime}, which shows the
numerically calculated populations of each level of the \JOneZero
system as a function of time when the field is modulated as in
Eqs. (\ref{eq:PMfield}) and (\ref{eq:PMphase}). The field has been
applied for a time of about $1000/\gamma$, sufficient in this case
to reach the quasi-steady state in which the atomic state
evolution is periodic. The settling time depends on the initial
state of the atom, on the laser intensity, and on the modulation
rate.  The time evolution of the state populations seen in
Figure~\ref{fig:j10PolModVsTime} displays oscillations at the
harmonics of $\delta_{\text{PEM}}$ imposed on the field by the
modulation.

\begin{figure}
\includegraphics{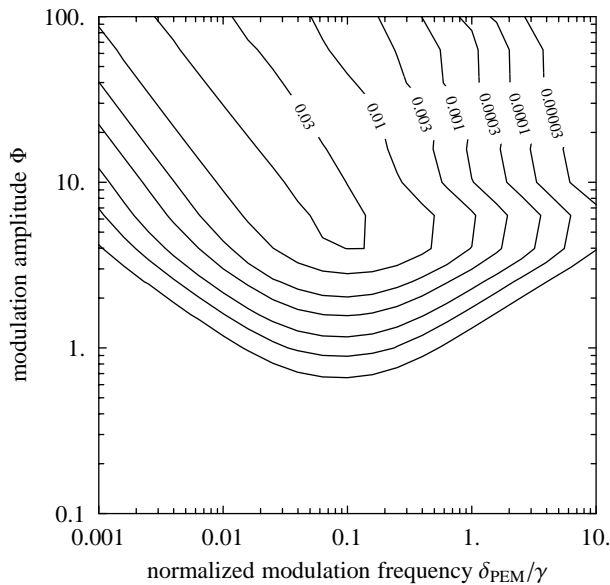}
\caption{Excited state population $P_f$ in the \JOneZero
transition as a function of normalized modulation frequency
$\delta_{\text{PEM}} / \gamma$ and modulation amplitude $\Phi$ for
the field given by Eqs.~(\ref{eq:PMfield}) and (\ref{eq:PMphase}).
Laser intensity and detuning as for
Fig.~\ref{fig:j10PolModVsTime}.} \label{fig:j10PolModAve}
\end{figure}

Figure~\ref{fig:j10PolModAve} shows the numerically calculated
population $P_f$ of the excited state (averaged over time $2\pi /
\delta_{\text{PEM}}$ in the quasi-steady state regime) as a
function of $\delta_{\text{PEM}}/\gamma$ and the phase modulation
amplitude $\Phi$. The Rabi frequencies and detunings are the same
as for Fig.~\ref{fig:j10vsAngle}, in which the dark states were
destabilized with a magnetic field. A comparison of the two graphs
shows that the two techniques can be similarly efficient at
destabilizing dark states. Figure~\ref{fig:j10PolModAve} shows
that the optimum modulation frequency is $\delta_{\text{PEM}} \sim
0.1 \gamma$ for small modulation amplitudes $\Phi \lesssim \pi$,
moving to lower values as $\Phi$ increases above $\pi$ and the
amplitude of the sidebands increases. The excited state population
is small when $\Phi \delta_{\text{PEM}} \ll \Omega$, because the
atom then adiabatically follows the slowly evolving dark state. It
is also small when $\Phi \delta_{\text{PEM}} \gg \gamma$ because
then much of the power of the modulated field is at frequencies
that are far from resonance.

In many experimental situations it may not be possible to
propagate the modulated and static linearly-polarized beams at
right angles \cite{Berkeland:1998}.  We have therefore repeated
the calculation of Fig.~\ref{fig:j10PolModAve} with the
unmodulated beam polarized at angle $\arccos \frac{1}{\sqrt{3}}
\sim 63^\circ$ to the propagation direction of the modulated beam.
The time evolution of the three polarization components of the
field is no longer completely orthogonal, and so the excited state
population is reduced for all modulation frequencies, in this case
by about a factor of three. However, the modulation frequency at
which the excited state population is maximized does not change
appreciably, nor does the frequency bandwidth of the atomic
response to the modulation.

In correspondence with the magnetic field case discussed in the
previous section, we find from our simulations that the linewidth
is large when the laser intensity is high ($\Omega > \gamma$),
when the modulation significantly broadens the laser frequency
spectrum ($\Phi \delta_{\text{PEM}}> \gamma$) and when the
evolution rate of the dark state is low ($\Phi \delta_{\text{PEM}}
\ll \Omega$).

Finally, we remark that although this second polarization
modulation scheme may not be the best approach for the \JOneZero
system because it needs a relatively expensive PEM, a variation of
it is probably the most appealing method for destabilizing dark
states in every other two-level system in zero magnetic field.  In
these systems, it is sufficient to use a field having only two
polarization components with different time dependences.  This can
be accomplished very simply by passing a single beam through an
electro-optic modulator (EOM), as we will discuss in the next
section.

\subsection{\SPD}
\label{subsec:SPD}

\begin{figure}
\includegraphics{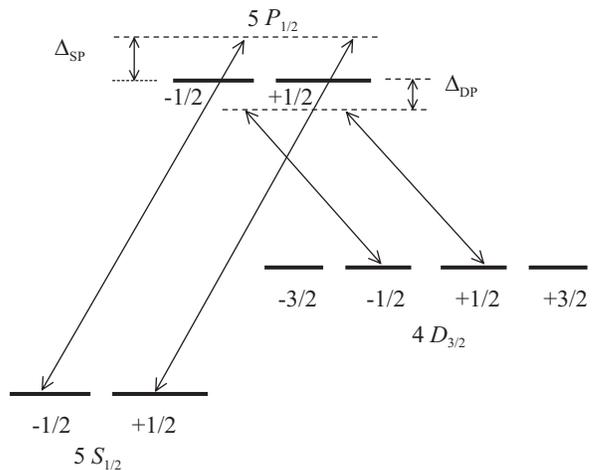}
\caption{Partial level diagram of $^{88}\text{Sr}^+$ showing
transitions driven by $\pi$-polarized laser light. This level
diagram also applies to $^{40}\text{Ca}^+$ and $^{138}\text{Ba}^+$
after respectively decrementing and incrementing the principal
quantum numbers by one.} \label{fig:Srlevels}
\end{figure}

In this section we consider the \SPD $\Lambda$-system, which
occurs in $^{40}\text{Ca}^+$ \cite{Knoop:1999, Urabe:1998,
Nagerl:1999, Hughes:1998, Block:1999}, $^{88}\text{Sr}^+$
\cite{Barwood:1998, Bernard:1999, Berkeland:1}, and
$^{138}\text{Ba}^+$\cite{DeVoe:1996, Raab:1998, Nagourney:1997,
Huesmann:1999} ions. Figure~\ref{fig:Srlevels} shows a partial
energy level diagram of these atoms. The $^2 P_{1/2}$ states decay
to both the $^2 S_{1/2}$ and the metastable $^2 D_{3/2}$ levels
with a branching ratio that favors the ${^2 P_{1/2}} \to {^2
S_{1/2}}$ decay (1:12 for Ca$^{+}$, 1:13 for Sr$^{+}$ and 1:2.7
for Ba$^{+}$). Because driving the \SP laser cooling transition
optically pumps the atom into the metastable $^2 D_{3/2}$ level, a
second "repumping" laser is tuned near resonance with the \DP
transition to pump the ion out of the $^2 D_{3/2}$ states. For
simplicity we assume that the $^2 D_{3/2}$ state is stable, which
is reasonable because the lifetime of this state is far greater
than any other time scale of the system. The rms Rabi frequencies
and detunings of the cooling and repumping lasers are denoted by
$\Omega_{\text{SP}}$, $\Delta_{\text{SP}}$, $\Omega_{\text{DP}}$,
and $\Delta_{\text{DP}}$ respectively.

Although the $J_i=\frac{1}{2} \leftrightarrow J_f=\frac{1}{2}$
cooling transition does not have a dark state, the
$J_i=\frac{3}{2} \leftrightarrow J_f=\frac{1}{2}$ repumping
transition does (Table~\ref{tab:darkTable2}). In the following
sections we discuss methods for destabilizing this dark state.  As
above, we consider first destabilizion with a magnetic field,
followed by a discussion of the polarization modulation technique.
We will also show how coherent population trapping in dark states
affects the lineshape of the cooling transition.

Convenient analytic solutions of the equation of motion for the
density matrix in this complex system are not possible, and so all
results presented for the \SPD system are based on numerical
solutions.

\subsubsection{Destabilization with a magnetic field}
\label{subsubsec:SPDBField}

\begin{figure}
\includegraphics{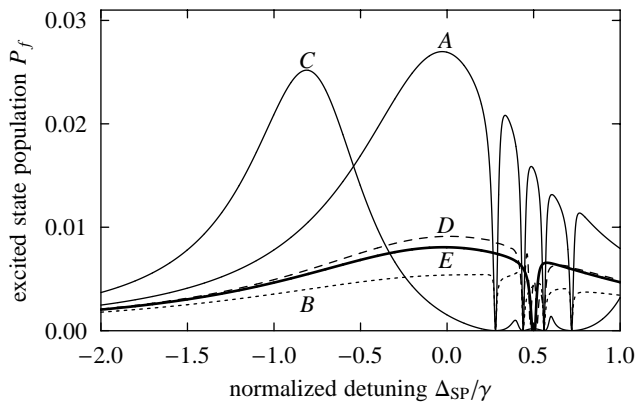}
\caption{Population of the $^2 P_{1/2}$ level of the
$^{88}\text{Sr}^+$ \SPD system in a magnetic field as a function
of normalized laser detuning $\Delta_{\text{SP}} / \gamma$.  Both
lasers are linearly polarized at angle $\theta_{\text{BE}}$ to the
magnetic field direction,
$\Omega_{\text{SP}}=\frac{\sqrt{2}}{5}\gamma$, and
$\Delta_{\text{DP}}=+\gamma/2$. Curve \textit{A}, and all other
curves unless noted otherwise: $\theta_{\text{BE}}=90^{\circ}$,
$\Omega_{\text{DP}}= \frac{\sqrt{2}}{5} \gamma$,
$\delta_B=0.1\gamma$. Curve \textit{B}: $\Omega_{\text{DP}}=
\frac{\sqrt{2}}{20} \gamma$. Curve \textit{C}:
$\Omega_{\text{DP}}= 2 \sqrt{2} \gamma$. Curve \textit{D}:
$\theta_{\text{BE}}=10^{\circ}$. Curve \textit{E}:
$\delta_B=0.003\gamma$.} \label{fig:fig7}
\end{figure}

When a magnetic field is applied to destabilize dark states in
this eight-state system, the transition lineshapes develop rich
structure.  This can be seen in Fig.~\ref{fig:fig7}, which shows
the cooling transition lineshape for different values of magnetic
field strength, laser polarization angle, and repumping laser
intensity. The structure seen around
 $\Delta_{\text{SP}}=+\gamma/2$ in each graph in the figure is due to
coherent population trapping in superpositions of $^2 S_{1/2}$ and
$^2 D_{3/2}$ states. These dark resonances have already been
studied in several experiments with trapped ions \cite{Janik:1985,
Siemers:1992}.

Figure~\ref{fig:fig7} shows that the cooling transition lineshape
is sensitive to the magnetic field, to the polarization angle, and
to the intensity of the repumping laser. In cases where the
evolution rate of the dark state is low, either because the the
magnetic field is small (curve $E$) or the polarization angle is
small (curve $D$), the resonance is broadened, for the reasons
discussed above in Sec.~\ref{subsubsec:J10BField}.  If the
repumping laser intensity is high (curve $C$), then the dark
resonances are power-broadened and the \SP transition displays a
substantial ac Stark shift.  On the other hand, when the repumping
laser intensity is low (curve $B$), the resonance is again
broadened.

The broadening seen in curve $B$ is closely related to the
broadening already encountered Sec.~\ref{subsubsec:J10BField} in
the regime where the dark state evolves too slowly. Its origin can
be understood here most simply in terms of rate equations
\cite{Janik:1985}. Figure~\ref{fig:3Lambda}(b) shows a simplified
version of the system, in which two lasers drive a simple $\Lambda
$-system in which the excited state $|f\rangle$ decays to the
ground states $|i\rangle$ and $|d\rangle$ with branching ratio
$(1-\alpha):\alpha$. The $|i\rangle \leftrightarrow |f\rangle$
transition has excitation rate $R_{\text{if}}$ and the $|d\rangle
\leftrightarrow |f\rangle$ transition has rate $R_{\text{df}}$.
The steady-state rate equations for this system are easily solved
to give the following expression for the excited state population
$P_f$:
\begin{equation}
\label{eq:LambdaSteadyState}
 \frac{1}{P_f} = 3 +
 \frac{(1-\alpha)\gamma}{R_{\text{if}}}+\frac{\alpha\gamma}{R_{\text{df}}}.
\end{equation}
When the rate on the $|d\rangle \leftrightarrow |f\rangle$
transition is low, the last term in this equation dominates, so
that the excited state population is insensitive to changes in the
rate of the $|i\rangle \leftrightarrow |f\rangle$ transition,
which in turn means that this transition is broadened. Physically,
the insensitivity arises because under these conditions most of
the population in the system is in state $|d\rangle$.  Increasing
the rate on the $|i\rangle \leftrightarrow |f\rangle$ transition
then has only a very small effect on the excited state population
because almost all of the population removed from state
$|i\rangle$ ends up in state $|d\rangle$.  There is an obvious
connection here to the picture for the broadening given above in
terms of quantum jumps. It follows from
Eq.~(\ref{eq:LambdaSteadyState}) that to avoid broadening the
cooling transition, the transition rates must be such that $\alpha
(1-\alpha) R_d > R_i$. In terms of the Rabi frequencies for the
\SPD system, this condition becomes $\Omega_{\text{DP}}^2 > \alpha
\Omega_{\text{SP}}^2$ in the limit $\alpha \ll 1$. Another
constraint on the intensities arises from the need to avoid
excessive power-broadening of the dark resonances by the repumping
laser, which translates into the restriction $\Omega_{\text{DP}} <
\gamma/2$.

\begin{figure}
\includegraphics{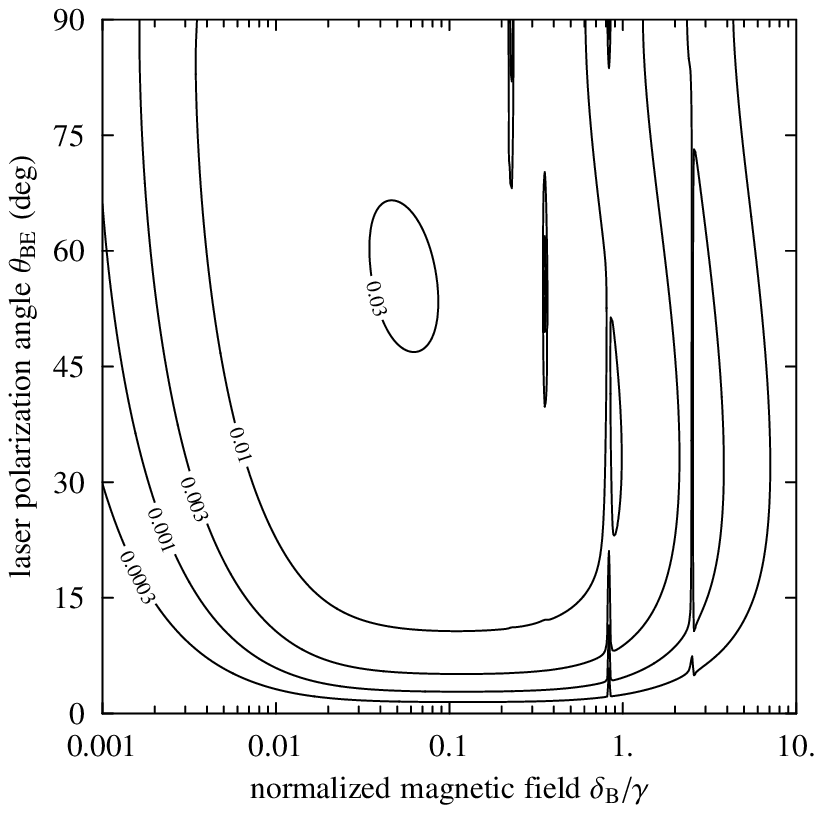}
\caption{Population of the $^2 P_{1/2}$ level of the
$^{88}\text{Sr}^+$  \SPD system as a function of magnetic field
strength and of the angle between the magnetic field and the
polarization vectors of the two laser fields. The rms Rabi
frequencies are $\Omega_{\text{SP}} = \Omega_{\text{DP}} =
\frac{\sqrt{2}}{5}\gamma$, and the detunings are
$\Delta_{\text{DP}} = +\gamma/2$ and $\Delta_{\text{SP}} = 0$. The
narrow vertical features are dips due to dark resonances (see also
 the thin curve in Fig.~\ref{fig:fig9}.)} \label{fig:fig8}
\end{figure}

\begin{figure}
\includegraphics{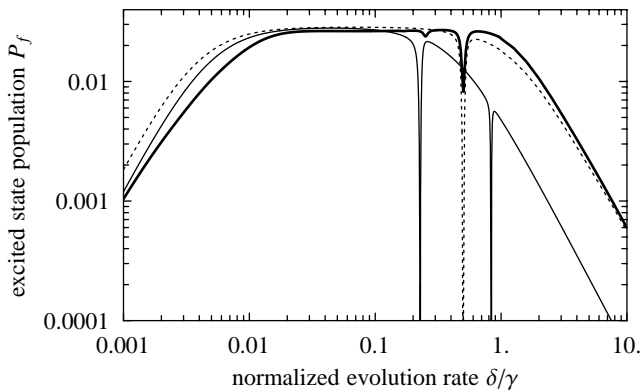}
\caption{Excited state population as a function of dark state
evolution rate $\delta$ for the $^{88}\text{Sr}^+$ \SPD system.
Thin curve: magnetic field applied at $\theta_{\text{BE}}=
90^{\circ}$, and $\delta=\delta_B$.  Thick curve: EOM polarization
modulation as in Eqs.~(\ref{eq:PM_EOM}) and (\ref{eq:PM_phase})
with $\Phi =\pi$, and $\delta=\delta_{\text{EOM}}$. Dashed curve:
AOM polarization modulation as in Eq.~(\ref{eq:PM_AOM}) with
$\delta_{\text{AOM+}}=-\delta_{\text{AOM-}}$, and
$\delta=\delta_{\text{AOM+}}$.  Laser intensities and detunings
are as for Fig.~\ref{fig:fig8}} \label{fig:fig9}
\end{figure}

While the structure seen in Fig.~\ref{fig:fig7} makes it difficult
to produce meaningful graphs of the linewidth (as in
Fig.~\ref{fig:j10WidthAndHeight}(b)), it is still straightforward
to plot the excited state population as a function of the
polarization angle $\theta_{\text{BE}}$ and magnetic field
strength (Fig.~\ref{fig:fig8}).  The laser frequencies and
intensities used here have been chosen to keep the dark resonances
on the blue side of the laser cooling transition and to avoid
power broadening. The graph is similar to the \JOneZero case
(Fig.~\ref{fig:j10vsAngle}), with two important differences.
First, the peak is broader in both angle and in magnetic field
strength.  The reduced sensitivity to the magnetic field arises
because this system does not pump into dark states as quickly as
the \JOneZero system, since most excited state decays are to the
$^2 S_{1/2}$ states and the repumping laser is detuned from
resonance. It follows that the system can tolerate slower dark
state evolution rates without adversely affecting the excited
state population. The second difference is the set of narrow
vertical dips, which are due to coherent population trapping in
dark states. These dips can be seen more clearly in the thin curve
in Figure~\ref{fig:fig9}, which shows the excited state population
as a function of the magnetic field for the same conditions as
those of Fig.~\ref{fig:fig8} and with $\theta_{\text{BE}}=
90^{\circ}$ (a convenient choice). The excited state population
vanishes for two magnetic fields. First, when $\delta_B =
\frac{5}{22}\gamma \approx 0.23 \gamma$ the
$|^2S_{1/2},m_i=+1/2\rangle$ and $|^2 D_{3/2}$,$m_i=-3/2\rangle$
states are Zeeman shifted into Raman resonance, which forms a
stable dark superposition of these two states. The same is true at
$\delta_B = \frac{5}{6}\gamma \approx 0.83 \gamma$, where the dark
state is composed of the $|^2S_{1/2},m_i=+1/2\rangle$ and
$|^2D_{3/2},m_i=+1/2\rangle$ states.  We note that for the
realistic Rabi frequencies used here, the optimum field strength,
corresponding to $\delta_B \sim 0.05\gamma$, is well removed from
the dark resonances.

We conclude this section with a discussion of the optimum
parameter values for destabilizing dark states with a magnetic
field without excessively broadening the transition.  The
intensity of the cooling laser should be set to drive the cooling
transition as hard as possible without power broadening it, which
will be the case if $\Omega_{\text{SP}} \sim \gamma/3$. Setting
the repumping laser intensity so that $\Omega_{\text{DP}} \sim
\gamma/3$, will make the repumping efficient without excessively
power-broadening the overall transition or the dark resonances.
Since the branching ratio $\alpha \ll 1$, this choice also avoids
excess broadening of the cooling transition from the mechanism
discussed after Eq.~(\ref{eq:LambdaSteadyState}).  The detuning of
the repumping laser used in Fig.~\ref{fig:fig7},
$\Delta_{\text{DP}}=+\gamma/2$, was chosen because it ensures that
the laser still drives the repumping transition efficiently while
keeping the dark resonances far from the red side of the cooling
transition, where the cooling laser is tuned during Doppler
cooling.  The magnetic field strength should be chosen so that
$0.01\gamma < \delta_B < 0.1\gamma$ (the upper limit is less than
the value of $~\gamma$ implied by Fig.~\ref{fig:fig8} because we
also wish to confine the dark resonances to a region of width
$<\gamma$ to keep them all on the red side of the resonance.)
Finally, although we have performed the calculations for light
which is linearly polarized perpendicular to the magnetic field,
Fig.~\ref{fig:fig8} shows that any polarization angle greater than
$15^{\circ}$ works well. In fact, our simulations show that the
resonance curve is not changed significantly even if the laser
polarizations are perpendicular to each other or if the repumping
laser is circularly polarized.

\subsubsection{Destabilization with polarization modulation}
\label{subsubsec:SPDPolMod}

We now consider destabilizing dark states in the $^2 D_{3/2}$
level by modulating the repumping laser polarization. In this
system, as in every other except the \JOneZero system, it is
sufficient to use a field having only two polarization components
with different time dependences. Perhaps the simplest method of
achieving this is to pass a single laser beam through an
electro-optic modulator (EOM) acting as a variable waveplate to
produce the field
\begin{equation}
\label{eq:PM_EOM}
 \left(\begin{array}{c}
 E_{-1} \\
 E_{0}  \\
 E_{+1} \\
\end{array} \right) =
\frac{E_{EOM}}{\sqrt{2}} \left( \begin{array}{c}
 1 + i e^{-i \varphi(t)} \\
 0                       \\
 1 - i e^{-i \varphi(t)} \\
\end{array} \right),
\end{equation}
where we have again defined the quantization axis to be parallel
to the propagation direction of the beam.  The retardation
$\varphi(t)$ is similar to that of Eq.~(\ref{eq:PMphase}),
\begin{equation}
\label{eq:PM_phase}
 \varphi(t) = \frac{1}{2}\Phi \left[1 - \cos
(\delta_{EOM}t) \right],
\end{equation}
with $\delta_{\text{EOM}}$ being the EOM drive frequency. The
thick curve in Figure~\ref{fig:fig9} shows the time-averaged
excited state population when the polarization of the repumping
laser is modulated in this way. We see that the excited state
population is reduced for certain values of $\delta_{\text{EOM}}$.
This is because the Fourier transform of the field of
Eq.~(\ref{eq:PM_EOM}), like that of Eq.~(\ref{eq:PMfield}),
contains harmonics of $\delta_{\text{EOM}}$ up to harmonic number
$\sim 2 \pi / \Phi$ when the modulation index exceeds one (so that
the effective dark state evolution rate in this high modulation
index regime is $\delta \sim \Phi \delta_{\text{EOM}}$).  The
excited state population will be reduced when one of these
sidebands connects the $^2 D_{3/2}$ and $^2 S_{1/2}$ states in
Raman resonance.  In this case, where the modulation index is
$\sim 1$, two such dips are visible. The excited state population
does not vanish completely in these dips because the other
frequency components present in the field still act to destabilize
the dark state.

Another easily-realized modulation scheme gives the $\sigma^+$ and
$\sigma^-$ polarization components different frequencies. The beam
can be split into right- and left-hand circular polarization
components separately shifted in frequency with two AOM's, to
create the field
\begin{equation}
\label{eq:PM_AOM} \left(\begin{array}{c}
 E_{-1} \\
 E_{0}  \\
 E_{+1} \\
\end{array} \right) =
\left( \begin{array}{c}
 E_{\sigma^+} e^{-i\delta_{\text{AOM-}} t }\\
 0       \\
 E_{\sigma^-} e^{-i\delta_{\text{AOM+}} t }\\
\end{array} \right).
\end{equation}
Although it is not necessary that
$\delta_{\text{AOM+}}=-\delta_{\text{AOM-}}$, this symmetrical
modulation most effectively destabilizes the dark state. The
dashed curve in Figure~\ref{fig:fig9} shows the excited state
population for this type of modulation. The result is similar to
the magnetic field and EOM methods, except that there is only a
single dark resonance, at
$\delta_{\text{AOM}}=-\Delta_{\text{DP}}$.

The similarity of the three curves in Figs.~\ref{fig:fig9} means
that the discussion in the previous section of optimum parameter
values for the magnetic field method applies also to the
polarization modulation case, with the appropriate dark state
evolution rate ($\Phi \delta_{\text{EOM}}$ or
$\delta_{\text{AOM}}$) replacing $\delta_B$.

\subsubsection{Non-zero laser bandwidth}
\label{subsubsec:BroadLaser}

In this section we consider what happens when the short-term
linewidth $\delta \omega_L$ of either laser is larger than the
decay rate $\gamma$.  This situation can be incorporated into the
simulation by selectively increasing the decay rate of the optical
coherences on the relevant transitions. The dark superpositions of
$^2 S_{1/2}$ and $^2 D_{3/2}$ states are then unstable, so that
the depth of the dark resonances is reduced, which in turn
simplifies the destabilization problem. This point is of some
practical importance because the repumping transition in
$^{88}\text{Sr}^+$ is often driven with a multi-mode fiber laser
whose linewidth is many times $\gamma$.  We find in this case that
the line shapes of both transitions remain symmetric and
structureless as long as $\Omega_{\text{DP}} \ll \delta \omega_L$
(where here $\Omega_{\text{DP}}$ is the Rabi frequency for a
single-mode repumping laser of the same intensity).  The intensity
of the broadband repumping laser intensity should then simply be
increased, subject to this limit, to maximize the excited state
population.  The rate at which the atom is pumped into the $^2
D_{3/2}$ dark states is then determined by the intensity of the
cooling laser, and so the other parameters should be set as
discussed above:  the magnetic field or polarization modulation
frequency should make the state evolution rate $\delta$ less than
the linewidth $\gamma$ and comparable to the cooling transition
Rabi frequency $\Omega_{\text{SP}}$.

The multi-mode nature of the fibre laser also makes possible a
particularly convenient form of polarization modulation on the \DP
transition \cite{Sinclair:2001}. A superposition of two fiber
laser beams having different polarization vectors produces a field
whose direction changes on the time scale corresponding to the
laser mode spacing.  This interval can be comparable to $\gamma$,
with the result that the simple two-beam arrangement can
effectively destabilize the $^2 D_{3/2}$ dark states (see
Fig.~\ref{fig:fig9}) without magnetic fields or modulators.

\subsection{Large angular momentum systems}
\label{subsec:HigherAM}

The responses of the \JOneZero and \SPD systems to the
destabilization techniques discussed above are strikingly similar.
This observation leads us in this section to consider atoms with
higher angular momentum, to demonstrate that the behavior seen in
those systems is for the most part quite general. For reasons of
computational simplicity, we consider only the destabilization of
dark states with a magnetic field, noting that we have seen in the
previous sections that the response to polarization modulation is
expected to be similar. We consider here a two-level atom with an
$S_{J_i}$ ground state and a $P_{J_f}$ excited state. We assume
the nuclear spin is zero and vary the electronic spin $S$ to
increase the total angular momentum, so these generic atoms have
no hyperfine structure.

\begin{figure*}
\includegraphics{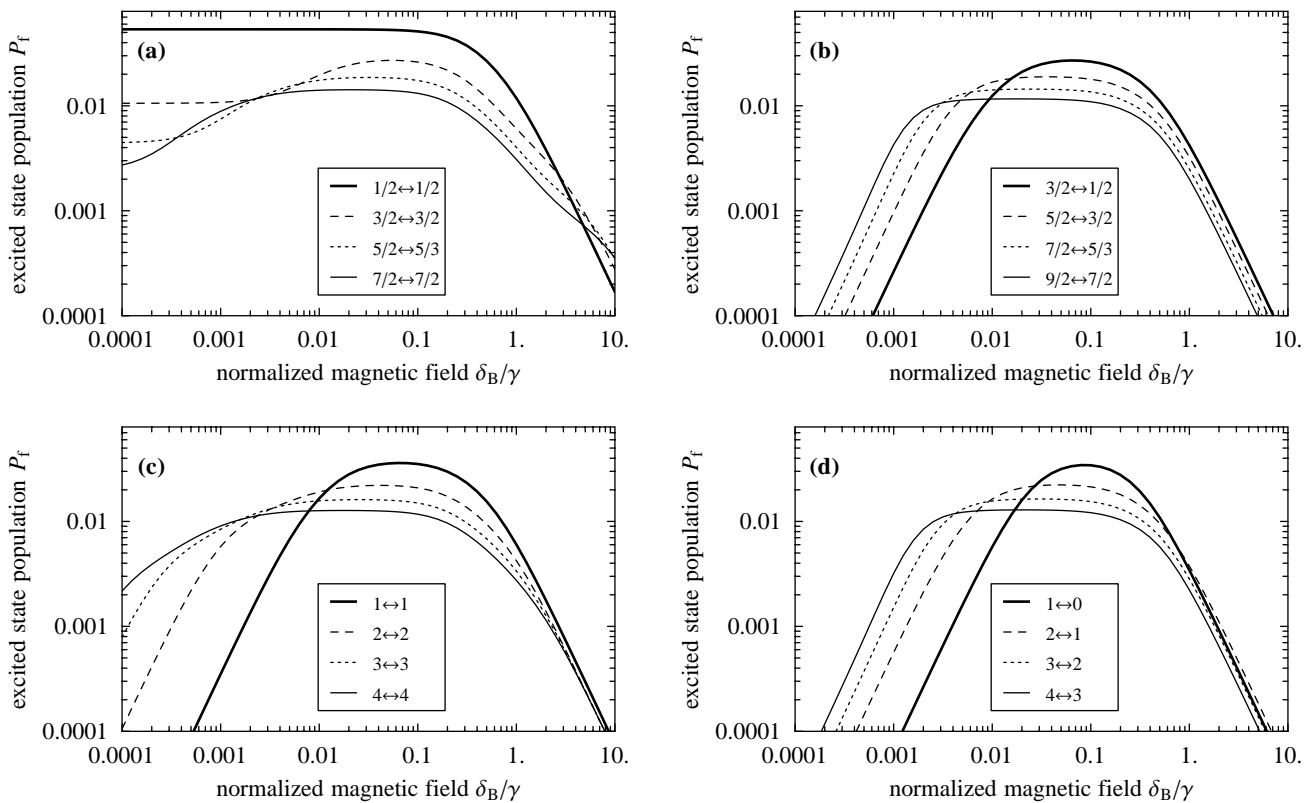}
\caption{Excited state populations as a function of magnetic field
for $J_i \leftrightarrow J_i -1$ and $J_i \leftrightarrow J_i -1$
transitions. In all cases, the rms Rabi frequency is $\Omega =
\frac{\sqrt{3}}{5}\gamma$, the polarization angle
$\theta_{\text{BE}} = \arccos \frac{1}{\sqrt{3}}$, and the laser
detuning $\Delta=0$.} \label{fig:fig10}
\end{figure*}

Figure~\ref{fig:fig10} shows the numerically calculated total
population $P_f$ of the $P_{J_f}$ levels as a function of magnetic
field, with linear laser polarization and polarization angle
$\theta_{\text{BE}} = \arccos \frac{1}{\sqrt{3}}$. The laser is
tuned to resonance, and its intensity gives an rms Rabi frequency
$\Omega = \frac{\sqrt{3}}{5}\gamma$, as in
Fig.~\ref{fig:j10vsAngle}. The graphs in this figure all show that
the excited state population is large when $\delta_B \sim
\Omega/4$. In graph (a) of Fig.~\ref{fig:fig10}, the excited state
populations do not vanish when $\delta_B$ approaches zero, since
the half-integer $J_i \leftrightarrow J_f$ transitions have no
dark state. However, in these transitions the excited state
populations can still be increased by applying a magnetic field,
especially when $J$ is large. Curves (b) -- (d) display two simple
scaling laws: the excited state population falls off like 1/$B^2$
when $\delta_B \gg \gamma$ because the atomic transition becomes
Zeeman broadened, and it grows as $B^2$ in the regime $\delta_B
\ll \gamma$ as the pumping out of the dark state(s) becomes more
efficient. These three curves also show that as $J$ increases, the
region in which the excited state population is large expands to
include smaller values of $\delta_B$. This is because increasing
the number of states in the system increases the time needed for
optical pumping into a dark state, and so the dark state evolution
rate $\delta$ can be smaller. Similarly, had the laser been
detuned from resonance then the excited state population would be
relatively constant over a range that includes smaller values of
$\delta$, because the atom would pump into the dark states less
rapidly. We find that the calculations resulting in
Fig.~\ref{fig:j10WidthAndHeight} for the width and excited state
population in the \JOneZero transition give very similar results
(not shown here) for these other two-level transitions.

In real atoms, large values of the total angular momentum are
usually due to the presence of nuclear spin. In this case the
$g$-factors of the atomic levels involved will be different than
those used in Fig.~\ref{fig:fig10}, which mostly results in a
simple shift along the $x$-axis of the appropriate curve. More
importantly, when the nuclear spin is not zero the atom can be
optically pumped into hyperfine levels which do not absorb light
from the laser field. This leads to situations which are similar
to those shown in Fig.~\ref{fig:3Lambda}, with a repumping
mechanism being needed to return the atom to the light state.
There are several ways of accomplishing this. Corresponding to
Fig.~\ref{fig:3Lambda}(a), an rf or microwave field can drive
transitions between the ground state hyperfine levels, or a second
laser field can pump the atom out of the extra ground state
hyperfine level through an auxiliary excited state. Corresponding
to Fig.~\ref{fig:3Lambda}(b), a repumping laser field can couple
the dark state to the same excited states as the main laser. For
all of these cases the conclusions are the same as in the previous
sections: the rate at which the atom is pumped out of the dark
hyperfine level must be as large as possible without exaggerating
coherence effects such as dark resonances. This means that the
polarization of the radiation driving the transition out of the
hyperfine states must be such that these hyperfine states do not
have a stationary dark state, and that the intensity of this
radiation must be great enough to drive the transition strongly.
If these conditions are not met then the width of the primary
transition will be broadened and the maximum scattering rate will
be reduced.

It it also useful to consider the opposite limit in which optical
pumping into dark hyperfine states has little effect on the
scattering process because the decay rate into these states is
sufficiently slow. In this case it may be possible to detect the
scattered photons with a time resolution which is much smaller
than the decay time to the dark hyperfine level. The dark periods
following decay into the dark state can then be selectively
neglected, with the result that the lineshape of the strong
transition will not be broadened, in contrast to the situation
discussed in Sec.~\ref{subsubsec:J10BField} where only the average
scattering rate was detected. Another consequence of weak coupling
to a dark hyperfine state is that the Doppler-cooled atom may be
able to reach equilibrium long before it decays into the dark
state \cite{Berkeland:1998}. The ultimate temperature of the atom
will then be the same as if the atom had only two levels (as long
as the atom is not heated while it is in the dark state).

\section{Discussion and conclusion}
\label{sec:Discussion}

In this paper we have discussed how the accumulation of atomic
population in dark states can be prevented by either applying a
static magnetic field or by modulating the polarization of the
driving laser. We have also considered the effect of these
destabilization techniques on transition lineshapes.  The magnetic
field technique is simple, but the more complex polarization
modulation method has the important advantage of leaving the
atomic energy levels unperturbed.  Several different atomic
systems were analyzed, and their responses to both techniques were
found to be quite similar (compare Figs.~\ref{fig:j10PolModAve},
\ref{fig:fig9}, and \ref{fig:fig10}). This universal behavior
arises because the evolution is always governed by the same
fundamental parameters: the state evolution rate $\delta$ (given
by the Zeeman shift, the AOM splitting, or the highest sideband
frequency for the case of phase modulation), the Rabi frequency
$\Omega$, and the excited state decay rate $\gamma$.

For a given laser intensity, the excited state population and the
scattering rate are maximized by making $\delta$ comparable to
$\Omega$ (typically $\delta \sim \Omega/2$).  The excited state
population will be small if $\Omega \gg \delta$ because the atom
is then able to follow the evolving dark state adiabatically, and
it will be small if $\Omega \ll \delta$, either because the laser
intensity is low ($\Omega < \gamma$) or because the atom and the
laser are detuned ($\delta > \gamma$).  If the transition
linewidth is not important, then the scattering rate can be
maximized by making $\Omega$ (and $\delta$) larger than $\gamma$,
so that the transition is saturated. If the linewidth is important
(e.g., in laser cooling applications), then the choice $\Omega
\sim \gamma/3$ gives substantial excited state population without
excessive broadening.  The two regimes which give small excited
state population ($\Omega \gg \gamma$ and $\Omega \ll \gamma$)
also result in broad lineshapes. Fortunately, the evolution rate
which optimizes the excited state population also minimizes the
linewidth.

If the system has more than two levels, then these rules apply to
the extra transitions if they too are to remain narrow. However,
often only one transition (for example, a laser cooling
transition) must be narrow. In this case the extra transition
should be driven as hard as possible if the system cannot form
dark resonances. If the system can form dark superposition states
(e.g. the \SPD system), then the intensities of the lasers should
give Rabi frequencies such that $\Omega < \gamma / 2$, to keep the
dark resonances narrow. In addition, because the dark resonances
occur when lasers are equally detuned from resonance, the laser
frequencies can be set to keep them away from any region of
interest.

\begin{acknowledgments}
It is a pleasure to thank James Bergquist for useful discussions
and Richard Hughes for carefully reading this manuscript. This
work was supported in part by ARDA under NSA Economy Act Order
MOD708600 and by the U.K. EPSRC.
\end{acknowledgments}


\begin{thebibliography}{20}
\bibitem{Orriols:1979} G. Orriols, Nuovo Cimento \textbf{53}, 1 (1979).
\bibitem{Arimondo:1976} E Arimondo and G. Orriols, Lett. Nuovo Cimento \textbf{17}, 333 (1976).
\bibitem{Gray:1978} H. R. Gray, R. M. Whitley, and C. R. Stroud, Jr., Opt. Lett. \textbf{3}, 218 (1978).
\bibitem{For:1998} For a recent review, see K. Bergmann, H. Theuer, and B. W. Shore, Rev. Mod. Phys. \textbf{70}, 1003 (1998).
\bibitem{Aspect:1988} A. Aspect, E. Arimondo, R. Kaiser, N. Vansteenkiste, and C. Cohen-Tannoudji, Phys. Rev. Lett. \textbf{61}, 826 (1988).
\bibitem{Harris:1989} S. E. Harris, Phys. Rev. Lett. \textbf{62}, 1033 (1989).
\bibitem{obald:1989} G. Th\'{e}obald, N. Dimarcq, V. Giordano and P. C\'{e}rez, Opt. Commun. \textbf{71}, 256 (1989).
\bibitem{Wineland:1975} D. J. Wineland and H. Dehmelt, Bull. Am. Phys. Soc. \textbf{20}, 637 (1975).
\bibitem{Hansch:1975} T. W. H\"{a}nsch and A. L. Schawlow, Opt. Commun. \textbf{13}, 68 (1975).
\bibitem{Arimondo:1996} E. Arimondo, in \textit{Progress in Optics XXXV}, edited by E. Wolf (N. Holland Pub. Co, Amsterdam, 1996) p. 257.
\bibitem{Bruce:1990}Bruce W. Shore, \textit{The Theory of Coherent Atomic Excitation}, (Wiley, New York, 1990).
\bibitem{Janik:1985} G. Janik, W. Nagourney, and H. Dehmelt, J. Opt. Soc. Am. B \textbf{2}, 1251-1257 (1985).
\bibitem{Janik:1984}G. R. Janik, Ph.D. thesis, University of Washington -- University Micro-films International, 1984.
\bibitem{Jon:1994} Jon H. Shirley and R. E. Drullinger, in \textit{Conference on Precision Electromagnetic Measurements Digest}, (IEEE, New York, 1994), p. 150.
\bibitem{Barwood:1998}G. P. Barwood, P. Gill, G. Huang, H. A. Klein, and W. R. C. Rowley, Opt. Commun. \textbf{151}, 50 (1998).
\bibitem{Berkeland:1998}D. J. Berkeland, J. D. Miller, J. C. Bergquist, W. M. Itano, and D. J. Wineland, Phys. Rev. Lett. \textbf{80}, 2089 (1998).
\bibitem{Weissbluth:1978} M. Weissbluth, \textit{Atoms and Molecules} (Academic Press, New York, 1978).
\bibitem{Edmonds:1960}A. R. Edmonds, \textit{Angular Momentum in Quantum Mechanics} (Princeton University Press, Princeton, NJ, 1960).
\bibitem{Alan:2001} Alan A. Madej and John E. Bernard, in \textit{Frequency Measurement and Control: Advanced Techniques and Future Trends}, and references therein, edited by A. Luitey (Springer-Verlag, New York, 2001).
\bibitem{Peter:1997}Peter T. H. Fisk, Rep. Prog. Phys. \textbf{60}, 761 (1997), and references therein.
\bibitem{Cardimona:1982} D. A. Cardimona, M. G. Raymer, and C. R. Stroud, Jr., J. Phys. B \textbf{15}, 55 (1982).
\bibitem{Many:1} Many of the symbolic and numerical calculations presented in this paper were performed with \textsc{Mathematica}$^{{\rm T}{\rm M}}$, a product of Wolfram Research, Inc., 100 Trade Center Drive, Champaign, IL, USA. Reference to this product does not imply an endorsement of the University of Sussex or Los Alamos National Laboratory.
\bibitem{Wayne:1982} Wayne M. Itano and D. J. Wineland, Phys. Rev. A \textbf{25}, 35 (1982).
\bibitem{Boshier:2000} M. G. Boshier, G. P. Barwood, G. Huang, and H. A. Klein, Appl. Phys. B \textbf{71}, 51 (2000).
\bibitem{Roberts:1999} M. Roberts, P. Taylor, S. V. Gateva-Kostova, R. B. M. Clarke, W. R. C. Rowley, and P. Gill, Phys. Rev. A \textbf{60}, 2867 (1999).
\bibitem{Enders:1993}V. Enders, Ph. Courteille, R. Huesmann, L. S. Ma, W. Neuhauser, R. Blatt, and P. E. Toschek, Europhysics Letters \textbf{24}, 325 (1993).
\bibitem{Fisk:1997}P. T. H. Fisk, M. J. Sellars, M. A. Lawn and C. Coles, IEEE Trans. Ultras. Ferr. Freq. Control \textbf{44}, 344 (1997).
\bibitem{Tamm:1995}C. Tamm, D. Schnier and A. Bauch, Appl. Phys. B \textbf{60}, 19 (1995).
\bibitem{Knoop:1999} M. Knoop, M. Vedel, M. Houssin, T. Schweizer, T. Pawletko, and F. Vedel, in \textit{Trapped Charged Particles and Fundamental Physics}, (AIP Conference Proceedings no. 457, 1999) p. 365.
\bibitem{Urabe:1998}S. Urabe, M. Watanabe, H. Imajo, K. Hayasaka, U. Tanaka, and R. Ohmukai, Appl. Phys. B \textbf{67}, 223 (1998).
\bibitem{Nagerl:1999}H. C. N\"{a}gerl, D. Leibfried, H. Rohde, G. Thalhammer, J. Eschner, F. Schmidt-Kaler, and R. Blatt, Phys. Rev. A \textbf{60}, 145 (1999).
\bibitem{Hughes:1998}R. J. Hughes and D. F. W. James, Forsch. Phys. \textbf{46}, 759 (1998).
\bibitem{Block:1999}M. Block, O. Rehm, P. Seibert, and G. Werth, Eur. Phys. J. D \textbf{7}, 461 (1999).
\bibitem{Bernard:1999}J. E. Bernard, A. A. Madej, L. Marmet, B. G. Whitford, K. J. Siemsen, and S. Cundy, Phys. Rev. Lett. \textbf{82}, 3228 (1999).
\bibitem{Berkeland:1}D. J. Berkeland ``A simple linear Paul trap for Sr$^+$'' (in preparation).
\bibitem{DeVoe:1996} R. G. DeVoe and R. G. Brewer, Phys. Rev. Lett. \textbf{76}, 2049 (1996).
\bibitem{Raab:1998}C. Raab, J. Bolle, H. Oberst, J. Eschner, F. Schmidt-Kaler, and R. Blatt, Appl. Phys. B \textbf{67}, 683 (1998).
\bibitem{Nagourney:1997}N. Yu, W. Nagourney, H. Dehmelt, Phys. Rev. Lett. \textbf{78}, 4898 (1997).
\bibitem{Huesmann:1999}R. Huesmann, C. Balzer, P. Courteille, W. Neuhauser, and P. E. Toschek, Phys. Rev. Lett. \textbf{82}, 1611 (1999).
\bibitem{Siemers:1992}I. Siemers, M. Schubert, R. Blatt, W. Neuhauser, and P. E. Toschek, Europhys. Lett. \textbf{18}, 139 (1992).
\bibitem{Sinclair:2001}A. G. Sinclair, M. A. Wilson, and P. Gill, Opt. Commun. \textbf{190}, 193 (2001).

\end{thebibliography}
\end{document}